%% file: jcap.tex
\documentclass[a4paper,11pt]{article}

\pdfoutput=1
\usepackage{jcappub}

\usepackage{paralist}
\usepackage{graphicx}
\usepackage{enumitem}
\usepackage{latexsym}
\usepackage{amsfonts}
\usepackage{amssymb}
\usepackage{xcolor}
\usepackage[export]{adjustbox}
\usepackage{amsmath}
\usepackage[thinlines]{easytable}
\usepackage{dcolumn}
\usepackage{verbatim}
\usepackage{float}
\usepackage{xspace}
\usepackage[normalem]{ulem}

\input{universalnewcommands.tex}

\newcommand{\DLens}{D_{\rm L}}
\newcommand{\DSource}{D_{\rm S}}

\newcommand{\tE}{t_{\rm E}}

\allowdisplaybreaks

\title{Dark Classification Matters: Searching for Primordial Black Holes with LSST}

\author[a]{Miguel Crispim Romao,}
\emailAdd{miguel.romao@durham.ac.uk}
\affiliation[a]{Institute for Particle Physics Phenomenology, Department of Physics, Durham University, Durham DH1 3LE, U.K.}
\author[a]{Djuna Croon,} 
\emailAdd{djuna.l.croon@durham.ac.uk}
\author[a]{Benedict Crossey,} \emailAdd{benedict.g.crossey@durham.ac.uk}
\author[b]{Daniel Godines} 
\emailAdd{godines@nmsu.edu}
\affiliation[b]{New Mexico State University, 1780 E University Ave, Las Cruces, NM 88003, USA.}

\date{\today}

\abstract{We present projected constraints on the abundance of primordial black holes (PBHs) as a constituent of dark matter, based on microlensing observations from the upcoming Legacy Survey of Space and Time (LSST) at the Vera C. Rubin Observatory. 
We use a catalogue of microlensing light curves simulated with Rubin Observatory’s OpSims to demonstrate that competitive constraints crucially rely on minimising the false positive rate (FPR) of the classification algorithm. We propose the Bayesian information criterion and a Boosted Decision Tree as effective discriminators and compare their derived efficiency and FPR to a more standard $\chi^2$-test.}

\begin{document}
\maketitle
\flushbottom

\section{Introduction}
This week, the Vera C. Rubin Observatory has sent ripples of excitement through the astrophysics community with the release of its ground-breaking first images, offering an impressive glimpse of what to expect over the next decade as it carries out the Legacy Survey of Space and Time (LSST). Thanks to its wide sky coverage, duration and deep-imaging capabilities, LSST will offer an unparalleled opportunity to search for gravitational microlensing events (e.g. \cite{LSSTProjection,Abrams+23}). Since microlensing is sensitive  to compact objects regardless of their luminosity, it provides a powerful, model-independent method to detect compact dark matter (DM) objects such as primordial black holes (PBHs) (see e.g. \cite{1986ApJ...304....1P,EROS-2:2006ryy,Mroz:2017mvf,MACHO:1998qtf,Macho:2000nvd,EROS-2:2006ryy,Niikura:2017zjd,Niikura:2019kqi,Mroz:2019xfm,LSSTDarkMatterGroup:2019mwo,Abrams:2020jvs,Winch:2020cju,Mroz:2024mse,Mroz:2024wia,Griest:2013aaa,Zumalacarregui:2017qqd,Oguri:2017ock,Blaineau:2022nhy,Green:2016xgy,Green:2017qoa,Green:2020jor,2024ApJ...961..179P,Kaczmarek:2024dmp,Fardeen:2023euf,Perkins:2025hfr,2019RNAAS...3...58L}).

A key challenge in realising this potential is the identification of microlensing events in LSST’s sparsely and irregularly sampled light curves. In previous work \cite{CrispimRomao:2025pyl}, a subset of the present authors introduced the use of the Isolation Forest algorithm for semi-supervised anomaly detection, which was shown to be suitable for near-real-time application. In this work, we instead demonstrate that 1-year of LSST data can be used to probe unconstrained regimes of the compact DM parameter space, with greater advances to be made with the full 10-year dataset. 
Using simulated microlensing events and the most up-to-date LSST observing strategy, we assess how effectively PBH-induced events can be identified using full LSST light curves, and in case of non-observation, robust constraints can be derived. 

LSST will probe over a billion stars, which has important implications for the required accuracy of the classifier.
To find competitive constraints, we find that minimising the false positive rate (FPR) of the classifier is crucial. Limited by the size of the simulated data set, we discuss how to reliably extrapolate the bayesian information criterion (BIC) and the boosted decision tree (BDT) output to define a derived efficiency with FPR $ < 10^{-7}$.

As we will demonstrate, with this methodology LSST is expected to be sensitive to PBHs in the mass range from $10^{-6}$ to tens of solar masses. Within just one year of observations, LSST is projected to improve upon existing galactic microlensing constraints by approximately half an order of magnitude.\footnote{We note that these results were obtained using a different treatment of the FPR. Given that the classification algorithms also differ and the full datasets are not publicly available, a direct replication of the analysis here is not feasible.} With the full anticipated 10-year survey duration, this sensitivity is expected to improve by an additional order of magnitude. Future foreground discrimination using multiband information may further strengthen the constraints on DM objects.

\section{Identifying microlensing at LSST}
\begin{figure}
    \centering
    \includegraphics[width=.8\linewidth]{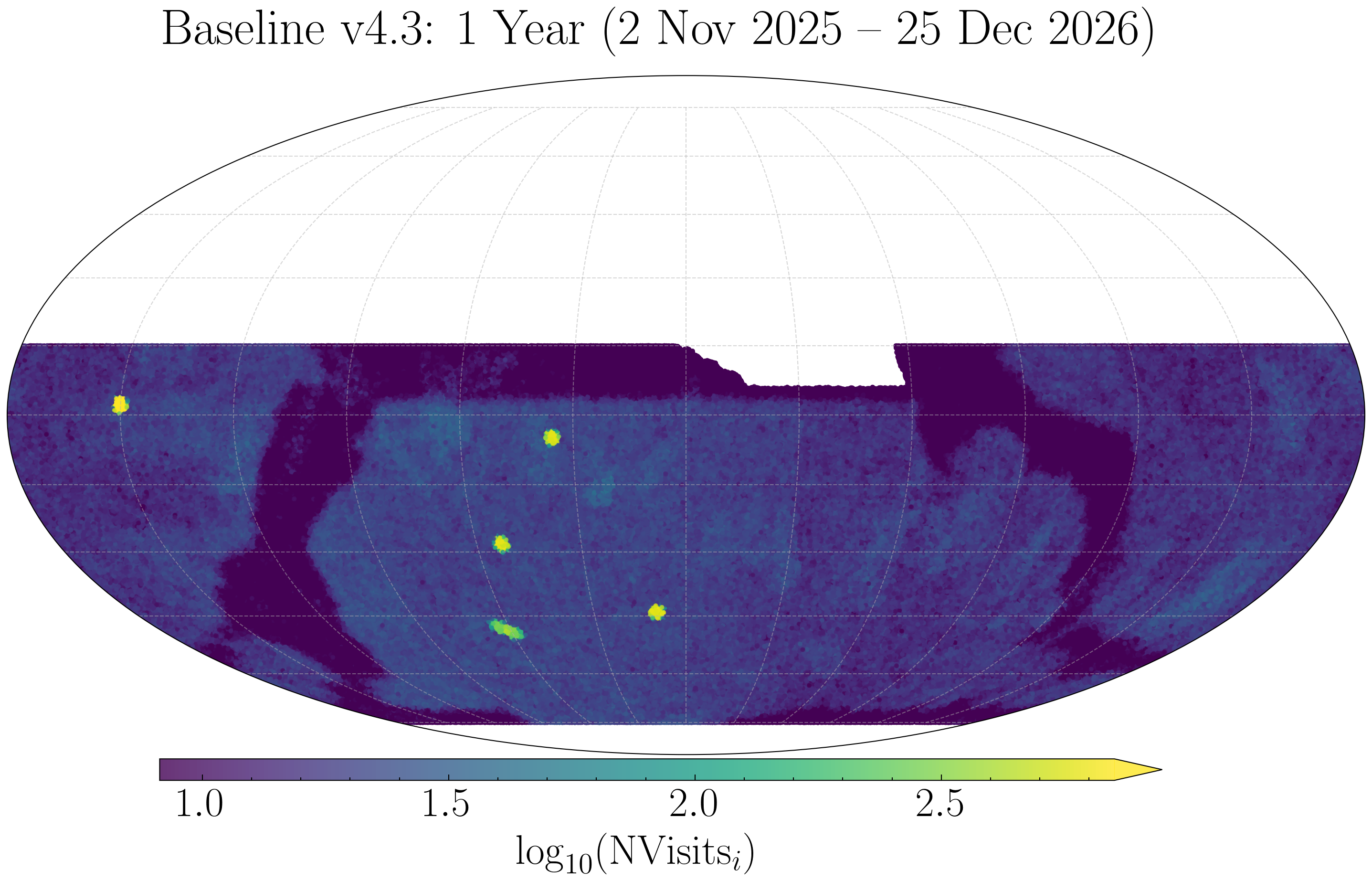}
    \caption{Skymap of the number of $i$-band visits (\texttt{NVisits$_i$}) during the first year of the LSST survey, simulated using the \texttt{baseline\_v4.3.1\_10yrs} strategy. The coverage reflects the depth variations across the sky due to the adopted survey cadence. We restrict the map to $-75^\circ\leq \delta \leq+15^\circ$ where the expected airmass is $\leq 1.4$. The plot is centered at right ascension $\alpha=0^\circ$ and declination $\delta=0^\circ$, with RA and declination lines marked every 15$^\circ$.}
    \label{fig:i-band-skymap}
\end{figure}

We use \texttt{rubin-sim} to simulate both signal-less (constant) light curves and point-source microlensing events based on the \texttt{baseline\_v4.3.1\_10yrs} survey strategy. This baseline uses the same strategy as the v4.0 simulations, which implement the final Phase 3 recommendations of the Survey Cadence Optimization Committee \citep{SCOC_Phase3}. The v4.3.1 cadence differs primarily in its updated survey start date and is used here to simulate the first year (up to 25 December 2026) of the LSST observations. We restrict our simulations to $i$-band visits, which offer greater depth than $g$-band in crowded fields where microlensing events are more likely to be detected. These observations are obtained using the two ``snap'' strategy (i.e., 2$\times$15s exposures). We draw sky coordinates uniformly within declinations $-75^\circ\leq \rm \delta \leq+15^\circ$, corresponding to regions expected to be observed with airmass $\leq 1.4$. A skymap of the number of $i$-band visits (\texttt{NVisits$_i$}) during the first survey year is shown in Fig.~\ref{fig:i-band-skymap}. 

The dataset consists of the light curves and derived time-series observables and statistics, which were obtained using \texttt{MicroLIA}~\cite{godines2019machine} and are the independent variables used to train the machine learning models. The dataset was divided into train, validation, and test sets with the training set used for training, the validation set for hyperparameter optimisation, and the test set for the analysis presented in this work. We restrict our study to the signal-less 
(Constant) and point-like lens microlensing (ML) classes. Both the training and validation sets are composed of $10^5$ Constant and ML classes, while the test set has $8.8\times 10^5$ Constant and $2.8\times 10^5$ ML light curves. The reason for the increased statistics of the test set will become apparent in the analysis, where a proper statistical description of tails of distributions is crucial to compute expected false positive rates. 

To simulate the ML class, consisting of point-like microlensing light curves, the theoretical magnification $A(t)$ is computed using three microlensing parameters: the lens Einstein crossing time, $\tE$, the lens minimal impact parameter, $u_0$, and the time of maximum magnification, $t_0$. The parameter $t_0$ is randomly drawn between the 1st and 99th percentiles of the observational window, with the window extended by half of $\tE$ on both sides to ensure sufficient sampling around the peak. This guarantees that a microlensing event occurs during the observational period, although it may not align with the observation cadence. Consequently, the $t_0$ distribution depends on $t_E$, and we note that while this choice ensures adequate sampling for training and evaluation, it may slightly overestimate the detection efficiency for long duration events whose peak would fall outside the observational window. The other microlensing parameters, $\tE$ and $u_0$, were sampled uniformly over the ranges presented in Table~\ref{tab:params} to cover a wide microlensing parameter space. 

The baseline magnitude is sampled uniformly between the LSST $i$-band saturation limit and its $5\sigma$ depth. A blending coefficient $g=F_B/F_S$, where $F_B$ and $F_S$ are the background and source fluxes respectively, is drawn uniformly between 0 and 1. The observed flux is then calculated as $F(t) = F_S(A(t) + g)$. 

\begin{table}
  \centering
  \begin{tabular}{cccc}
    \hline\hline
    parameter & min & max & spacing \\
    \hline
    $\tE$ (days) & 0 & 400 & linear \\
    $u_0$ & 0 & 3 & linear \\
    \hline\hline
  \end{tabular}
  \caption{Parameters used in the simulation of the microlensing class, ML. For more details see~\cite{CrispimRomao:2025pyl}.
  }
  \label{tab:params}
\end{table}

\begin{figure*}
    \centering
    \includegraphics[scale=0.4]{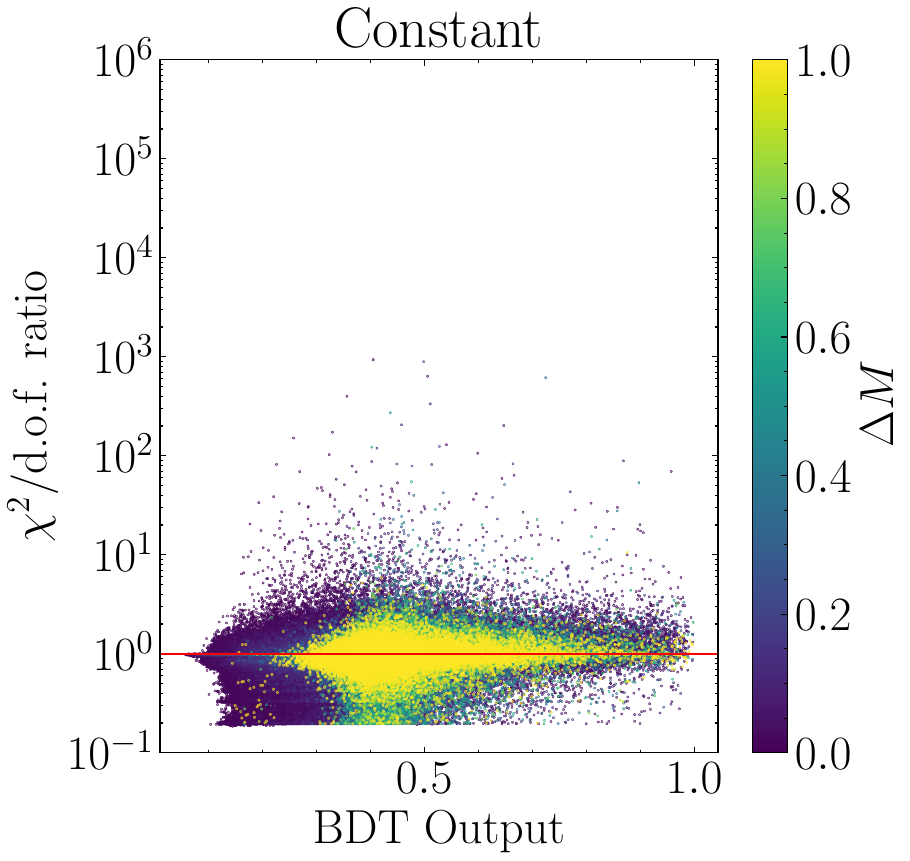}
    \includegraphics[scale=0.4]{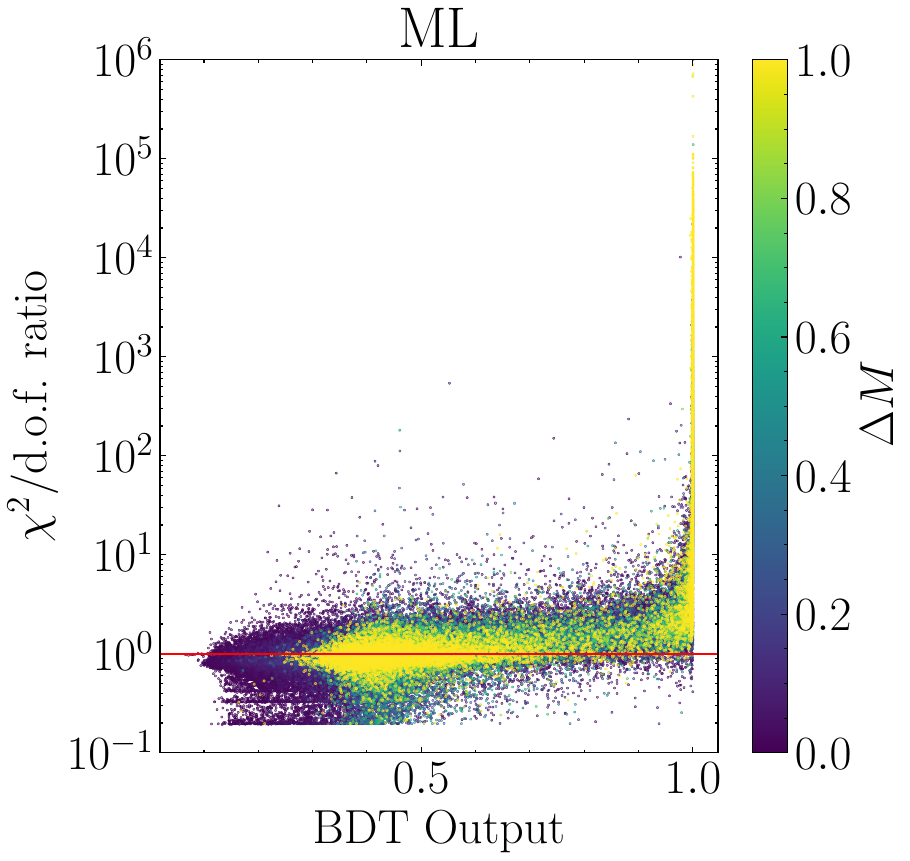}
    \includegraphics[scale=0.4]{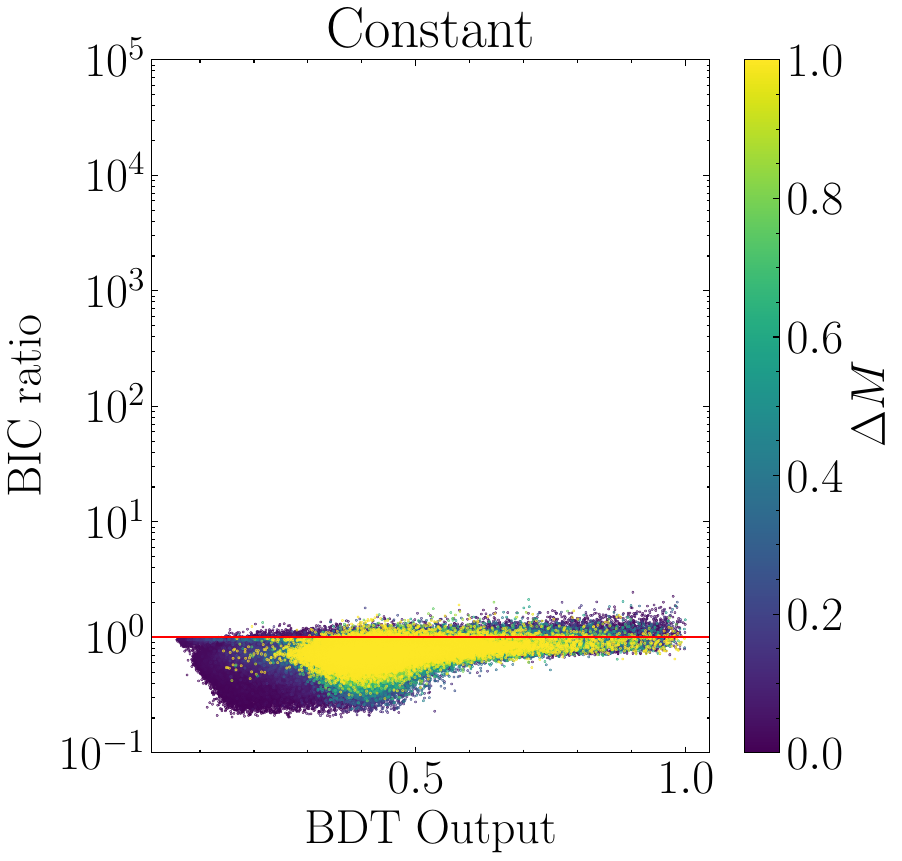}
    \includegraphics[scale=0.4]{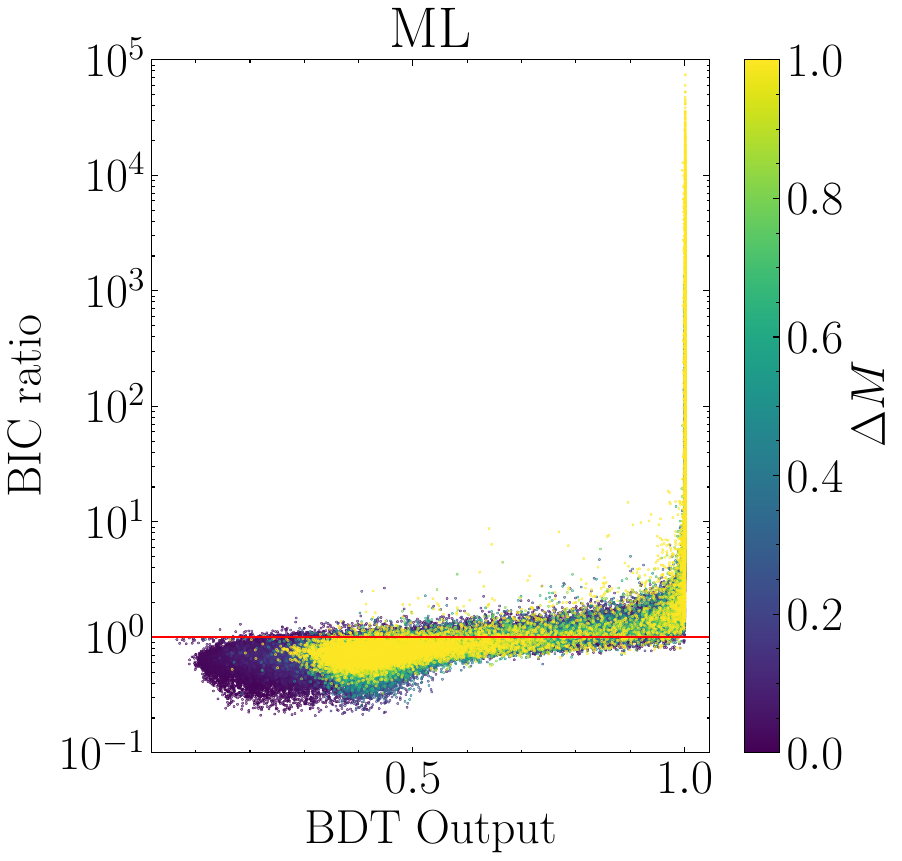}
    \caption{Classifier outcomes: $\chi^2 /\text{d.o.f. ratio}$, BDT output, and BIC ratio as defined above for the signal-less constant class on the left and microlensing events on the right. The magnitude difference in the event is indicated by the colour gradient.
    }\label{fig/classifierresults}
\end{figure*}

We now assess the sensitivity of
different classification techniques for detecting microlensing of light curves indicative of PBH events using simulated light curves for the LSST. 
Specifically, our analysis compares three types of classifiers: 
\begin{itemize}
    \item A supervised Histogram-based Gradient Boosting Classification Tree (which we will refer to simply as BDT for Boosted Decision Tree) trained to distinguish between the two classes of light curves;
    \item A simple ratio test between $\chi^2$-fits to ML and Constant light curves (with a cut on the inferred minimum impact parameter $u_0$);
    \begin{equation}\label{eq:chi2-ratio}
        \chi^2 /\text{d.o.f. ratio} \equiv 
        \frac{\chi^2_{\rm Const}/\nu_{\rm Const}}{\chi^2_{\rm ML}/\nu_{\rm ML}},
    \end{equation}
    where $\nu$ is the number of degrees of freedom in the models. For the Constant model, the only parameter is the source apparent baseline magnitude, hence $\nu_{\rm Const}=n - 1$, where $n$ is the number of data points in the light curve. For the ML model, we have five parameters: the source apparent baseline magnitude, the minimal impact parameter, the Einstein crossing time, the time of minimal impact parameter, and the blending ratio, and hence  $\nu_{\rm ML}=n - 5$.\footnote{In principle, one could perform a likelihood ratio test by computing $| {\chi^2_{\rm Const}/\nu_{\rm Const}}-{\chi^2_{\rm ML}/\nu_{\rm ML}}| $, which follows a $\chi^2_{\nu_{\rm ML}-\nu_{\rm Const}}$ distribution from which a $p$-value can be obtained to assess whether one rejects the simpler model, i.e. Constant. Unfortunately, for a large number of data points this always favours the more complex model, i.e. ML. This makes this approach unsuitable for our work where the number of data points varies greatly between light curves, and can be large. As we will see, the $\chi^2$/d.o.f. ratio is a more tractable quantity to separate both classes, but still fails to penalise the more complex model for almost all Constant light curves.    
    }
    \item The Bayesian Information Criterion (BIC). The BIC for a model $\mathcal{M}$ with $k$ free parameters fitted to $n$ data points is defined as
    \begin{equation}\label{eq:bic-ratio}
        \mathrm{BIC}(\mathcal{M}) \;=\; k\ln n \;-\; 2\ln\hat L,
    \end{equation}
    where $\hat L$ is the maximised likelihood of the model, which is effectively given by the $\chi^2$ as  $- 2 \ln \hat L = \chi^2$. 
    
    A difference of BIC between two fitted models of $2$ or greater is often taken as good evidence to prefer the model with lowest BIC. However, in our case we found that this was not a good rule-of-thumb, as it leads to too many Constant light curves having a good ML fit. Alternatively, we will look into the ratio of the BIC of both fits
    \begin{equation}
        \mathrm{BIC}\text{ ratio} = \frac{\mathrm{BIC}(\mathcal{M}_{\rm{Const}})}{\mathrm{BIC}(\mathcal{M}_{\rm{ML}})}=\frac{\chi^2_{\rm{Const}}+\ln n}{\chi^2_{\rm{ML}} + 5 \ln n} \ .
    \end{equation} 
    Comparing to Eq.~\eqref{eq:chi2-ratio}, we see that the BIC penalises more complex models proportionally to the (log) size of the data. This will prevent bias towards complex models unless the $\chi^2$ is significantly lower than the one for the simpler model.
\end{itemize}
The BDT was explicitly trained on labelled examples of Constant and ML light curves. Hyperparameter tuning for the BDT was performed and the details can be seen in Appendix~\ref{app:hp-tuning}.

We show the results of our classifier study in Fig.~\ref{fig/classifierresults}.
The results demonstrate that truly constant light curves (left panels) can still produce very large $\chi^2$ ratios when fit with the microlensing model. This is explained by a fit to the noise rather than true events. 
We also observe that higher magnitude differences, $\Delta M$, do not necessarily lead to higher $\chi^2$ ratios and, in fact, we can see that higher values of $\chi^2$ ratio are not associated with high brightness magnifications. This challenges the utility of $\chi^2$ as a good discriminant to filter out signal-less light curves. 
By contrast, genuine microlensing curves (right panels) show very large $\chi^2$ ratios 
associated with high event amplitude magnification, indicating a genuine improvement in fit. Nonetheless, here too we observe many light curves with high magnification that are not associated with high $\chi^2$ ratios.
The BIC ratio (bottom row) mirrors this trend but penalises the extra parameters more harshly, so they remain close to unity for noise‐dominated fits and rise sharply only when there is a true microlensing signature. 
Lastly, it is clear from the horizontal distribution of datapoints that BDT scores near $1$ can be reached for ML events, whilst the Constant light curves are assigned systematically lower values. 
In practice, the most competitive constraints on the number of events observed in the survey are obtained by selecting threshold cuts on these ratio statistics to reject the bulk of false positives while retaining a high fraction of true microlensing detections.

To derive constraints on the DM fraction, we need to calculate the number of expected events each 
discriminant (BDT output and BIC ratio)
can confidently identify in the survey. Two quantities are important: the expected false positive rate (FPR) and the accompanying efficiency, i.e. the true positive rate. 
While we do not know how many true positives to expect, we can set an expected number of false positives as the number of Constant light curves that we expect to prevail above a certain threshold cut on the chosen discriminant.
We define the FPR as
\begin{equation}
    \text{FPR} = \frac{N_{\rm FP}}{N_{\rm FP} + N_{\rm TP}}
\end{equation}
where $N_{\rm FP}$ ($N_{\rm TP}$) is the number of false (true) positives the discriminants
find in our test data. 
As we will see below, competitive constraints necessitate a very small FPR, below the inverse number of light curves in a simulated dataset of reasonable size.
A practical approach to set this decision threshold is to consider the finite nature of our data.
Unfortunately, this also means that in order to have complete control over the expected FPR we need to have an arbitrairly large dataset for the analysis. For example, in our case we can at most predict FPR$\simeq 1/ (8.8\times10^{5})\simeq 10^{-6}$ per star per year (as our data was simulated to cover one year of observation). If we take into account that the LSST is designed to look at $10^9$ stars, this will still leave $\mathcal{O}(10^3)$ expected false positives per year, which can severely impact the sensitivity.
Additionally, relying solely on the maximum output value from the Constant class light curves would be misleading as with infinite data the tail of the distribution for the Constant class may eventually reach values that one would expect only to see for ML light curves.

Instead, we adopt an alternative approach that provides some control over the tail of the discriminants distributions  of the Constant class, allowing us to model their asymptotic behaviour and extrapolating to lower FPR. To achieve this, we will focus on the BIC ratio and BDT output distributions. Looking at the bottom left panel of Fig.~\ref{fig/classifierresults} we see that the BIC ratio decays very quickly for the Constant class, and that the bulk of the BDT output is for very low values. While modelling the whole distribution is very (and unnecessarily) ambitious, we now ``zoom'' in to the tails composed of the $1\%$ highest values of the BIC ratio and BDT output, and model this part of the distribution instead.

\begin{figure*}
    \centering
    \includegraphics[scale=0.4]{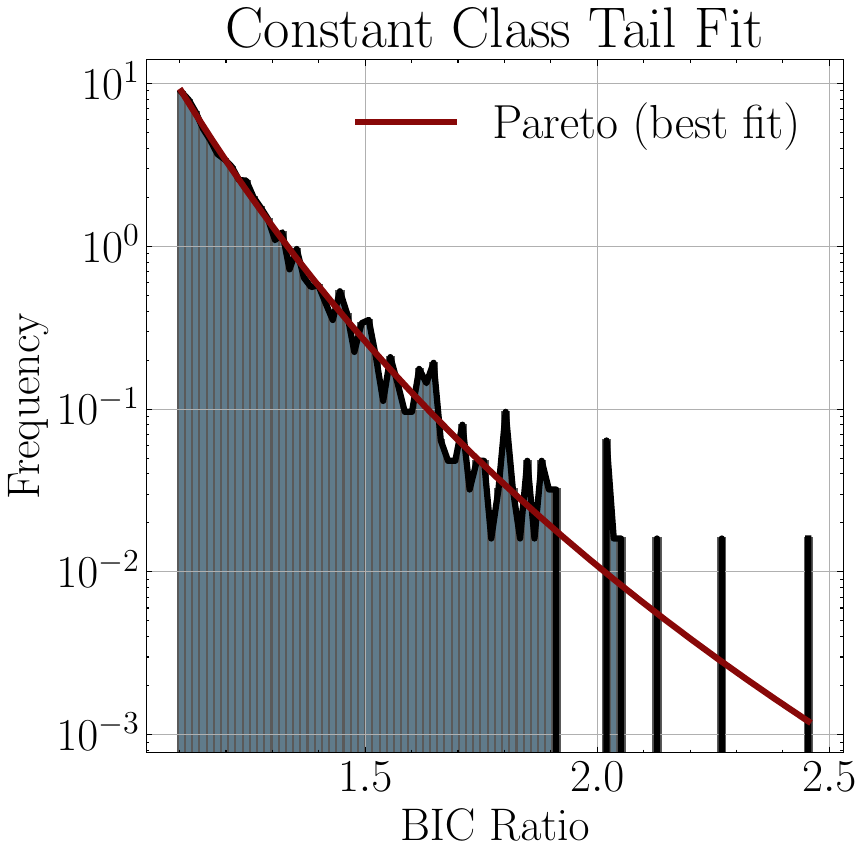}
    \includegraphics[scale=0.4]{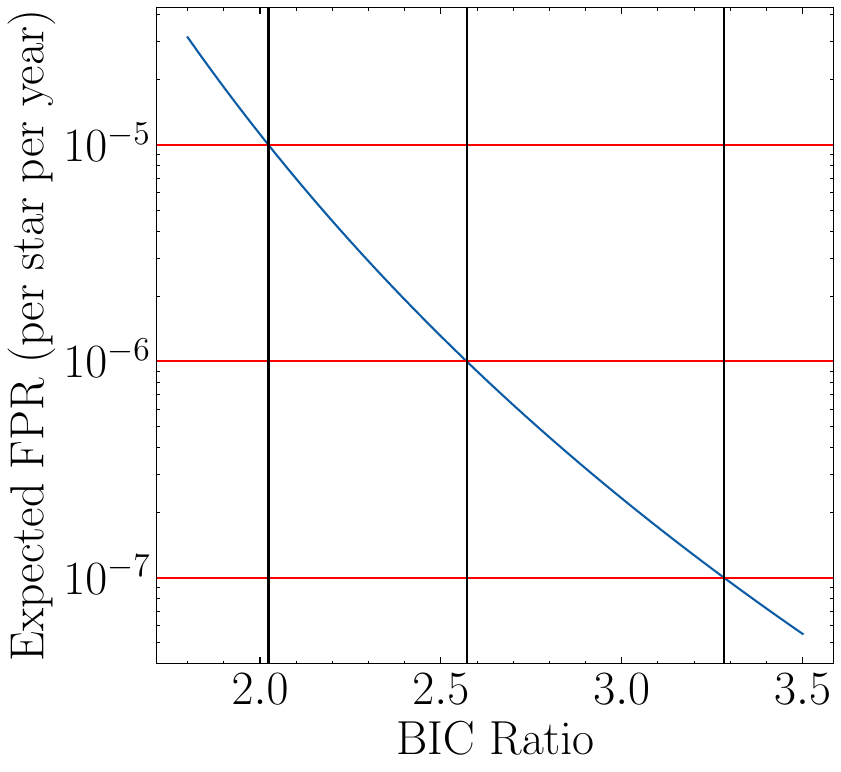}
    \includegraphics[scale=0.4]{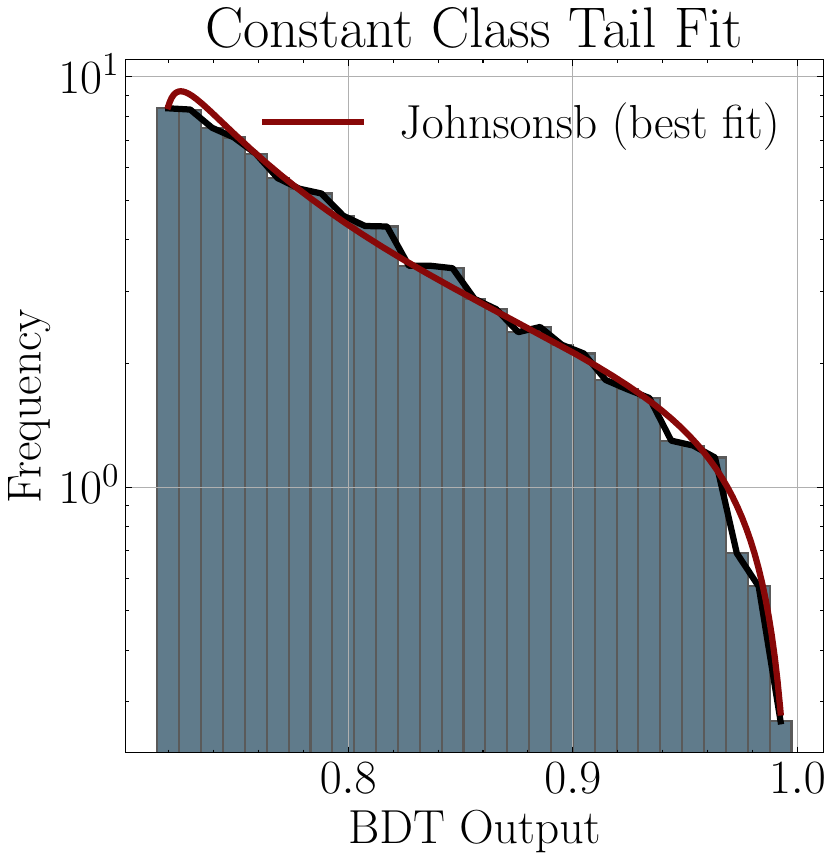}
    \includegraphics[scale=0.4]{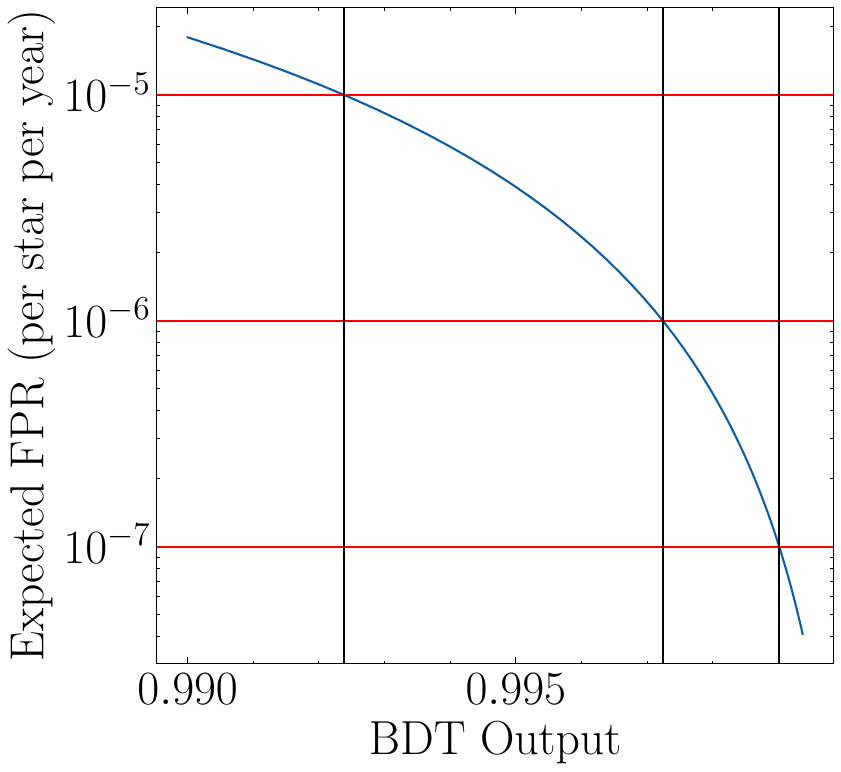}
    \caption{
    $99\%$ tail distributions of BIC ratio and BDT outputs for the constant class, and best fit closed-form distributions used to extrapolate to expected FPR $=10^{-7}$.}
    \label{fig/tail-fits}
\end{figure*}

To find the best distribution fit to the data, we use \href{https://github.com/erdogant/distfit}{\texttt{distfit}}, a \texttt{python} package that automatically fits all distributions available in \texttt{scipy} on a dataset and returns the best fit. In Fig.~\ref{fig/tail-fits} we show the results for this procedure, where we show the best fit over each tail as well as the predicted expected FPR over each discriminant. On the left panels, we present the fit to the BIC ratio for the Constant class tail. The distribution that better fits this tail is the Pareto distribution, which describes long tails of the form $b/x^b$.\footnote{The Pareto distribution probability density function is given by $f(x,b)=b/x^b$, which can be further rescaled and ``re-located'' by two parameters, $scale$ and $loc$ such that $f(x,b)\to f((x-loc)/scale,b)/scale$. The \texttt{distfit} best fit values at two significant digits were $b=8.8$, $scale=0.90$, $loc=0.19$.} We observe how the tail is very well described by the Pareto distribution, providing confidence on the extrapolated expected FPR per star per year presented in the second pane. In this pane, we show that we can expect an FPR per star per year of $10^{-7}$ for a BIC ratio cut of around $3.28$. We can now perform the same study for the BDT output, which has the added nuance that it is bounded. Fortunately, \texttt{distfit} also covers bounded distributions, and one can find that the Johnson $S_B$\footnote{The Johnson $S_B$ distribution is bounded and has a probability density function of the form $f(x,a,b)=(b/(x(1-x)))\phi(a+b\log(x/(1-x))$, where $\phi$ is the normal distribution probability density function, and it can be further rescaled and ``re-located'' by $f(x,b)\to f((x-loc)/scale,b)/scale$. The \texttt{distfit} best fit values at two significant digits were $a=0.73$,  $b=0.67$, $scale=0.29$, $loc=0.71$.} offers a good description of the sharp decay of BDT output values for the Constant class just below $1.0$, as can be seen in the third panel of Fig.~\ref{fig/tail-fits}. This results in the extrapolating expected FPR per star per year shown in the fourth panel, where we expect it to be $10^{-7}$ for a BDT cut at $0.999$.

In Table~\ref{tab:cuts} we list these two cuts, alongside the expected FPR per star per year as well as the global efficiency. We also provide two extra cut criteria that are inspired by cuts often considered in microlensing analysis: a good fit to the ML model embodied by $\chi^2$/d.o.f. ratio $>10$, and the demand that the \emph{fitted} minimal impact parameter is small, $\tilde u_0<1$. We see that the cuts using our discriminants, the BIC ratio and the BDT, are expected to have orders of magnitude lower FPR per star per year while providing higher efficiency. In the next section, we will compare the bounds on the dark matter fraction composed of PBHs obtained using these different cuts.
\begin{table}[t]
  \centering
  \begin{tabular}{cccc}
    \hline\hline
    cut & FPR & $\bar\epsilon$  \\
    \hline
    BIC ratio $>3.28$ & $10^{-7}$ & 0.38 \\
    BDT $>0.999$ & $10^{-7}$ & 0.34 \\
    $\chi^2$/d.o.f. ratio $>10$ & $3.5 \times 10^{-4}$ & $0.30$ \\
    $\chi^2$/d.o.f. ratio $>10$,  $\tilde u_0<1$ & $1.1 \times 10^{-4}$ & $0.20$ \\
    \hline\hline
  \end{tabular}
  \caption{
  Different cuts based on different discriminants and their respective false positive rate per star per year and global ML efficiency ($\bar\epsilon)$. $\tilde u_0$ refers to the $u_0$ estimate provided by the ML model fit to the light curve.
  }
  \label{tab:cuts}
\end{table}

\begin{figure}
    \centering
    \includegraphics[width=0.6\linewidth]{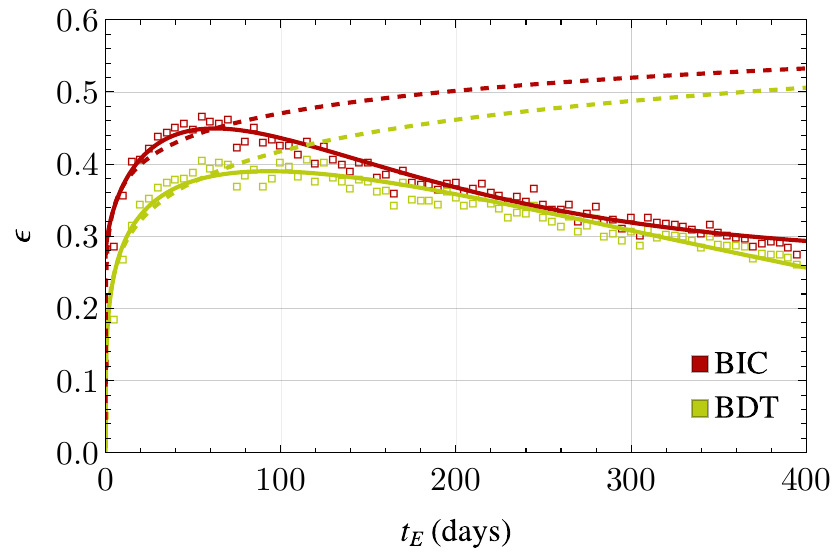}
    \caption{Efficiency after 1 year of data defined as the fraction of events for which the classifier output exceeds the cuts (BDT score and BIC as in table~\ref{tab:cuts}), for the two classifiers described in text. Here we binned the data linearly with $\Delta \tE = 5$ days. The dashed curves provide the fit of to the function \eqref{eq:efficiencyfunc} of the first $100$ days: it is seen that while this provides a reasonable fit for this period, it overestimates the efficiency for larger Einstein crossing times. The continuous curves provide the fit to \eqref{eq:efficiencyfunc2}.}
    \label{fig:efficiency_fit}
\end{figure}

Using the cuts described above, we can derive an effective sky-averaged efficiency as a function of Einstein crossing time $\tE$. 
We compute the percentage of simulated microlensing events, binned by event duration $ \tE $, for which the discriminant exceeds the cut. 
We used a bin width of $ \Delta \tE = 1.0 $, and each event is assigned to a bin according to $ \text{bin}(\tE) = \Delta \tE \cdot \lfloor \tE / \Delta \tE \rfloor $. For each unique bin, the total number of events is counted, along with the number for which the discriminant is greater than the threshold. The efficiency in each bin is then calculated as the percentage of events satisfying this condition, i.e., 
\begin{equation}
    \epsilon =  N_{>\text{cut}} / N_{\text{total}}.
\end{equation}
The result is displayed in Fig.~\ref{fig:efficiency_fit}, along with (in dashed curves) a fit to the analytic function suggested in \cite{Winch:2020cju}, 
\begin{equation}\label{eq:efficiencyfunc}
    \epsilon = \left(\left(\frac{\tE}{t_0}\right)^{-1/t_r}+1\right)^{-1}
\end{equation}
where we find $t_0 = 365 (194)$ days and $t_r = 3.90 (5.57)$ for the BDT (BIC) respectively. However, we caution that this function asymptotes to $\epsilon = 1$ at large Einstein crossing times, which is unrealistic and will overestimate the efficiency for large lens masses (indeed this bias may have affected the projections found for heavy mass objects in previous studies). Therefore, we suggest it is replaced by an efficiency derived from classifier results, or multiplied by a window function in $\tE$. Alternatively, we propose another analytic function which provides a better fit, 
\begin{equation}\label{eq:efficiencyfunc2}
    \epsilon = A \left(\frac{\tE}{\rm days}\right)^k e^{\lambda  (-\tE/\rm days)}+c
\end{equation}
where we find $A=0.22(2.8\times 10^{-2})$, $k=0.19(0.59)$, $c=-5.4\times10^{-2}(0.27)$, $\lambda=2.1(9.4)\times10^{-3}$ for the BDT (BIC) respectively.

\section{Projected constraints on PBHs}\label{sec:constraints}
We now proceed to estimate the event rate observed in LSST, based on the sensitivity described in the previous section. Our approach follows the methodology outlined in Ref.~\cite{Griest:1990vu,Croon:2020wpr}, which assumes the lenses follow the distribution of DM, 
\begin{equation}
  \rho(x) = f_{\rm DM} \rho_{\rm DM}(x),  
\end{equation}
where $f_{\rm DM}$ represents the fraction of dark matter composed of lensing objects, $x \equiv \DLens/\DSource$ is a normalised distance we integrate over to get constraints ($D_{S}$ is the distance to the source, $D_L$ is the distance to the lens), and $\rho_{\rm DM}$ is the overall dark matter density. Here we assume an isothermal profile, which simplifies the calculation of the lensing rate considerably as the circular speed of the lens $v_c$ is independent of distance $x$ (see \cite{Green:2025dut} for an analysis of the sensitivity of microlensing constraints to the density and velocity profiles).
It is given by
\begin{equation}
    \begin{split}
        \nn \rho_{\rm DM} (r) &= \frac{\rho_{\rm s}}{1+(r/r_{\rm s})^2}~,\\
        r &\equiv \sqrt{R^2_{\rm Sol} - 2 x R_{\rm Sol} \DSource \cos \ell \cos b + x^2 \DSource^2}~,
    \end{split}
\end{equation}
where $R_{\rm Sol} = 8.5\,\mathrm{kpc}$ is the Sun's galactocentric distance, $\rho_{\rm s} = 1.39\,\mathrm{GeV/cm^3}$ is the core density, and $r_{\rm s} = 4.38\,\mathrm{kpc}$ is the core radius (see~\cite{PPPPCookbook}). The coordinates $\ell$ and $b$ denote the galactic longitude and latitude of the source, respectively.

For a single source star observed over a unit time interval, the differential event rate as a function of $x$ and Einstein crossing timescale $t_{\rm E}$ is given by 
  \begin{equation}
    \frac{d\Gamma}{d\tE}=\frac{32 D_S u_T^4 \epsilon(\tE)}{\tE^4 v_c^2 M}\int_0^1 dx\rho(x)R_E^4(x)e^{-\frac{4R_E^2(x)u_T^2}{\tE^2 v_c^2}}
    \label{eq:dGammadt}
  \end{equation}
with Einstein Radius
  \begin{equation}
    R_E(x)=2\sqrt{GM D_S\,x(1-x)}
  \end{equation}
in natural units ($c=1$). Here $M$ is the lens mass, and $u_T$ is the threshold impact parameter $u\equiv b/R_E$ -- all smaller impact parameters produce a magnification above the threshold $T=1.34$ (thus for point-like lenses as considered here we have $u_T = 1$). 
We note that this differential event rate defines microlensing events as $u_0=1$, whilst our classifiers may be able to identify events with larger minimal impact parameters (with reduced efficiency). We thus expect our constraints to be conservative.
Eq.~\eqref{eq:dGammadt} is derived under the assumption that both the observer and the source remain stationary in the microlensing framework, which ensures an accuracy of approximately 10\% in event rate calculations~\cite{Niikura:2019kqi}. 
Additionally, it presumes that all lensing objects in the population share a uniform mass $M$, though extending this result to a broader mass distribution is straightforward (see e.g. \cite{Carr:2017jsz}, as well as a tool to carry out this generalisation in \href{https://github.com/SergioSevi/EDObounds}{EDObounds}~\cite{Croon:2024jhd}).

As discussed in the previous section, we employ an averaged detection efficiency function $\epsilon(\tE)$ that depends solely on the microlensing Einstein crossing timescale $\tE$. Because this function is derived using the LSST simulated events with the scheduled cadence, it inherently accounts for the increased detectability of longer-duration events. By using $\epsilon(\tE)$ in our rate calculation, we effectively average over the spatial variations—such as differences in stellar density, extinction, and observing conditions—that exist across the LSST footprint, thereby reducing the efficiency to a one-dimensional function of Einstein crossing timescale.

The total number of observed microlensing events is then determined by
\begin{equation}
N_{\rm events} = N_\star T_{\rm obs} \int^{t_{\rm E, max}}_{t_{\rm E, min}} d\tE \frac{d\Gamma}{d\tE}~,
\label{eq:Nevents}
\end{equation}
where $N_\star = 4 \times 10^9$ denotes the number of source stars the survey is sensitive to in a single visit in the i-band up to $\sim 23.9$  \cite{LSSTScience:2009jmu}, $T_{\rm obs}$ is the observation duration, and $t_{\rm E, min}$ ($t_{\rm E, max}$) corresponds to the minimum (maximum) timescale for an event to be observed within the survey.

In modelling the LSST microlensing foreground, we conservatively assume that main sequence stars produce light curves indistinguishable from point-like lenses, and that no spectroscopic or multi-band analysis is performed to distinguish these events. 
We use \verb|DarkDisk_Microlensing| \cite{Winch:2020cju} which we update to include a spherical Hernquist bulge next to the double‐exponential stellar disc to represent the baryonic lens population. The total stellar density is then defined as
\begin{equation}
     \rho_{\rm bar}(r,z) \;=\; \rho_{\rm disc}(r,z) + \rho_{\rm bulge}(R)\,, 
\end{equation}
with
\begin{equation}
\begin{split}
    \rho_{\rm disc}(r,z) &= A \exp\!\bigl[-(r - R_{0})/h_{r}\bigr] \exp\!\bigl[-\lvert z\rvert/h_{z}\bigr]\,, 
\\
  \rho_{\rm bulge}(R) &= \frac{M_{b}}{2\pi}\,\frac{r_{b}}{R\,(R + r_{b})^{3}}\,,   
\end{split}
\end{equation}
where \(A=0.04\,M_{\odot}\,\mathrm{pc}^{-3}\), \(h_{r}=3\,\mathrm{kpc}\), \(h_{z}=0.4\,\mathrm{kpc}\), \(R_{0}=8.2\,\mathrm{kpc}\), \(M_{b}=10^{10}\,M_{\odot}\) and \(r_{b}=1\,\mathrm{kpc}\). By sampling a Kroupa initial mass function and performing the microlensing integrals over lens distance and velocity, we generate per-star event‐rate histograms for a one-year LSST survey. 
In addition to main sequence stars, we further include a brown‐dwarf population by extending the mass function below the hydrogen‐burning limit. We adopt the Chabrier log‐normal IMF for sub‐stellar objects,
\[
  \xi_{\rm BD}(m)\;\propto\;\exp\!\Bigl[-\frac{\bigl(\log_{10}(m/m_{c})\bigr)^{2}}{2\sigma^{2}}\Bigr],
  \quad 0.01 \le \frac{m}{M_{\odot}} \le 0.08,
\]
with characteristic mass \(m_{c}=0.079\,M_{\odot}\) and width \(\sigma=0.69\). 
This allows us to compute the number of expected events per $\tE$-bin, now including bulge lenses and brown dwarfs, furnishing a baryonic foreground against which any DM microlensing signal may be contrasted.

\begin{figure*}
    \centering
    \includegraphics[width=0.95\linewidth]{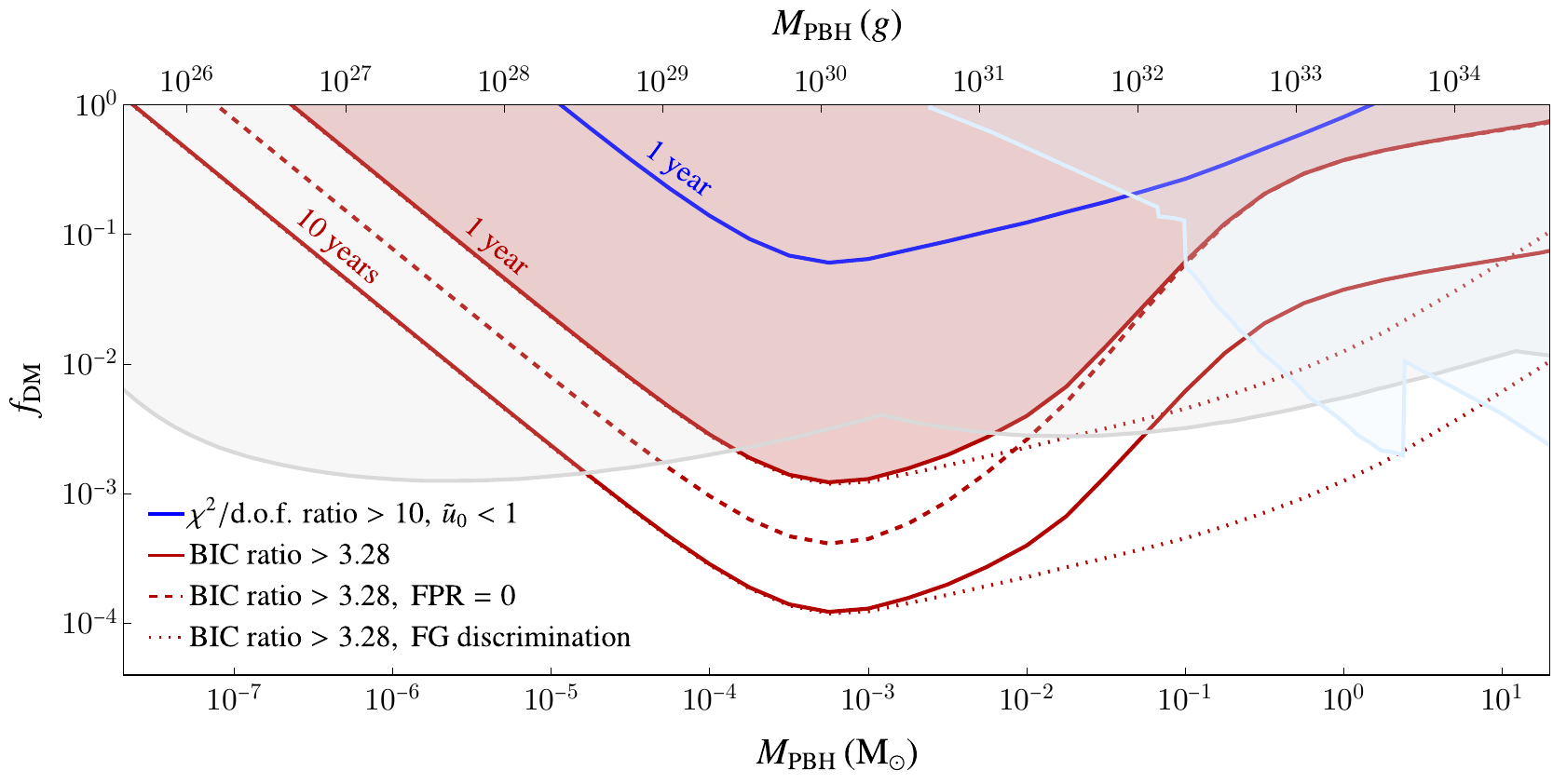}
    \caption{Projected constraints on the PBH parameter space from gravitational microlensing by the LSST. 
    We show the 1 year and the 10 year projections with filled and continuous lines respectively. In blue, we indicate the constraints obtained using the $\chi^2$/d.o.f. ratio test described in text along with a cut on the fitted $u_0$ parameter ($\tilde u_0$). In red, we show the constraints derived using the BIC with threshold cut as described; the BDT gives equivalent constraints. With the dashed line we indicate the constraints we would have derived had we ignored the FPR of our classifier. With the dotted lines we indicate the constraints if we assume the foreground can be distinguished using spectroscopic information. 
    In gray existing microlensing constraints, in light blue existing gravitational wave constraints. See text for further details. 
    }
    \label{fig:projections}
\end{figure*}

To obtain projected limits, we define $N^{\rm DM}_i$ as the number of PBH-induced events expected from \eqref{eq:Nevents}, and $N^{\rm FG,eff}_i = N^{\rm FG}_i +N^{\rm FP}_i $ as the effective foreground count for every $\tE$-bin $i$, where $N^{\rm FP}_i$ is the expected number of false positive events per bin. Then, we define the signal as $N^{\rm SIG}_i \equiv N^{\rm FG}_i + N^{\rm FP}_i+ N^{\rm DM}_i$. As in \cite{Croon:2020wpr}, we compute
\beq
\kappa = 2 \sum_{i = 1}^{N_{\rm bins}} \left[ N^{\rm FG,eff}_i - N^{\rm SIG}_i + N^{\rm SIG}_i \ln \frac{N^{\rm SIG}_i}{N^{\rm FG,eff}_i} \right]
\eeq
and obtain the 90\% \acro{c.l.} Poissonian limit by locating ($f_{\rm DM}, M$) for which $\kappa$ = 4.61~\cite{PDGStats}.

We present our results in Fig.~\ref{fig:projections}, alongside existing constraints on the dark matter parameter space from microlensing (grey, \cite{EROS-2:2006ryy,Mroz:2024mse,Mroz:2024wia}) and gravitational wave (non-)observations (light blue, \cite{Kavanagh:2018ggo,Chen:2019irf,LIGOScientific:2019kan,Nitz:2022ltl,Boybeyi:2024mhp}). These existing bounds were obtained from the \href{https://github.com/bradkav/PBHbounds/}{PBHbounds} repository \cite{kavanagh2019pbhbounds}.
Our 1-year exclusion projections are shown as filled red contours, based on the BIC-ratio cut listed in Table~\ref{tab:cuts}, corresponding to an extrapolated false positive rate (FPR) of $10^{-7}$. (We note that applying the equivalent BDT score cut yields nearly identical constraints, though we do not show it explicitly.)
We find that such a low FPR is required for our constraints to be competitive with existing bounds. This point is further illustrated by the projected constraints obtained using the $\chi^2$/d.o.f.\ cut from Table~\ref{tab:cuts} (shown in blue), which achieves comparable efficiency but results in an FPR roughly three orders of magnitude higher.
To emphasize the importance of careful FPR control, we also include a dashed line showing a naive projection assuming FPR$=0$.
For comparison, we show an optimistic scenario—where foregrounds can be effectively discriminated—using dotted lines. As expected, this has the largest impact at higher lens masses, where foreground contamination (e.g., from brown dwarfs and main-sequence stars) becomes more significant.
To estimate the projected constraints after 10 years of survey time, we apply a simple rescaling of Eq.~\eqref{eq:Nevents}. While the longer light curves will introduce slight changes in the detection efficiency at large Einstein crossing times, the resulting changes primarily affect the high-mass end of the parameter space, where our projections are already subdominant to existing microlensing constraints.

\section{Discussion}
In this work, we have compared the performance of several discriminants to identify microlensing events from point-like DM objects (such as primordial black holes and main sequence stars in the Galactic bulge and disk) from signal-less light curves simulated for the LSST survey. Specifically, we derived cuts 
in the BIC-ratio and the output of a boosted decision tree to minimise the false positive rate, and from this
derived a detection efficiency $\epsilon(\tE)$ which we present in Fig.~\ref{fig:efficiency_fit}. Based on this efficiency, we calculated the number of expected microlensing events during one and ten years of the survey, and used this to find projected constraints on the PBH parameter space. We present these projections in Fig.~\ref{fig:projections}.

A central focus of this work has been the careful treatment of the classifier’s FPR, which is crucial for translating microlensing detections into robust constraints on compact DM objects. 
In particular, we note that an incomplete treatment of the classifier FPR may lead to overoptimistic results, and may explain differences with prior studies.
We find that competitive bounds on the PBH abundance require extremely low FPRs -- on the order of $10^{-7}$ per star per year -- due to the sheer number of stars observed by LSST. Achieving such control is not feasible through brute-force simulation alone. We therefore modelled the high-end tails of our classification discriminants (BDT and BIC ratio) using closed-form probability distributions, allowing us to extrapolate FPRs well below the resolution of the test set.
This extrapolation enabled a principled selection of classifier thresholds and allowed us to quantify the global efficiency for detecting true microlensing events as a function of Einstein crossing time $\tE$. Importantly, we assumed that false positives are uniformly distributed in $\tE$, which is a conservative assumption that avoids overfitting to specific signal-like timescales. In reality, we expect false positives to be more prevalent in particular $\tE$ ranges (e.g., due to sparse sampling), and a more realistic modeling of their $\tE$ distribution may sharpen future projections.

While the averaged efficiency $\epsilon(\tE)$ captures the global trend of detectability as a function of event duration, it neglects spatial dependencies that arise due to variations in LSST's coverage. In reality, the detection efficiency should also vary with the source location, reflecting field-to-field differences in cadence, blending, and local observing conditions. A more refined analysis would involve developing a spatially resolved efficiency model $\epsilon(\tE, \vec{x})$ that explicitly incorporates position-dependent parameters. Such an approach would require detailed simulations of the LSST survey that consider the local environment in each field, thereby providing a more accurate prediction of the microlensing event rate. 
Furthermore, future simulations can incorporate realistic priors for parameters, informed by tools such as \href{https://github.com/jluastro/PopSyCLE}{\texttt{PopSyCLE}}~\cite{2024ApJ...970..169P,2025ApJ...980..103A}. By doing so, we can integrate astrophysical priors into our classification framework, modelling the prior probability distribution of microlensing parameters within the training set.

Although the primary goal of this study is to estimate LSST’s projected sensitivity to DM microlensing, the classification framework developed here may be enhanced by real-time event detection. If microlensing events can be flagged in near-real-time—using anomaly detection operating on nightly data, as was suggested in \cite{CrispimRomao:2025pyl}, this would open the door to photometric or spectroscopic follow-up of candidate events. Such follow-up could help break degeneracies, enhance the efficiency whilst reducing the FPR, as well as reduce contamination from foregrounds. 

In this work we focused on point-like lenses as a subfraction of DM. Importantly, we did not consider other DM lens classes as in \cite{CrispimRomao:2025pyl,CrispimRomao:2024nbr}, which can feature distinct signatures that can aid classification. Thus, we expect the results in this work to be valid for compact DM objects with physical radius well below the Einstein radius $R < R_E$ -- in practise this implies $R \lesssim\mathcal{O}(\rm AU)$ for the range of masses we constrain \cite{Croon:2020wpr,Croon:2020ouk} . Future work will contain an extension to other DM objects.

\section*{Acknowledgements}
We note that authorship ordering on this work is alphabetical; all authors have made important contributions to this work.
  
We thank Natasha Abrams, David McKeen, Nirmal Raj, Jessica Lu and the LSST Discovery Alliance microlensing subgroup for useful discussions.~MCR, DC, and BC are supported by the STFC under Grant No.~ST/T001011/1.


\appendix

\section{Hyperparameter Tuning\label{app:hp-tuning}}

The BDT hyperparameters were optimised via a hyperparameter optimisation loop using \texttt{optuna}~\cite{optuna}. In the loop, each proposed hyperparameter combination is used to train the BDT and validated against the Area Under the Curve (AUC) of the Receiver Operator Characteristic (ROC) computed using the validation set. The best combination of hyperparameters is identified as the one yielding the highest ROC AUC on the validation set. The ranges of hyperparameters considered for optimisation and their final optimal values are listed in Table~\ref{tab:hyperparameters}.
\begin{table}[h]
	\centering
	\begin{tabular}{ccc}
		\hline\hline
		Hyperparameter               & Range           & Optimal Value \\
		\hline
		\texttt{learning\_rate}      & $[0.01,\ 1]$    & $0.018$       \\
		\texttt{max\_iter}           & $[100,\ 1000]$    & $421$         \\
		\texttt{max\_leaf\_nodes}    & $[10,\ 100]$     & $69$          \\
		\texttt{min\_samples\_leaf}  & $[20,\ 200]$     & $90$          \\
		\texttt{l2\_regularisation}  & $[0.1,\ 1]$  & $0.20$       \\
		\hline\hline
	\end{tabular}
	\caption{Hyperparameter ranges and optimal values found during the hyperparameter optimisation step.}
	\label{tab:hyperparameters}
\end{table}

\begin{figure}
    \centering
    \includegraphics[width=0.75\linewidth]{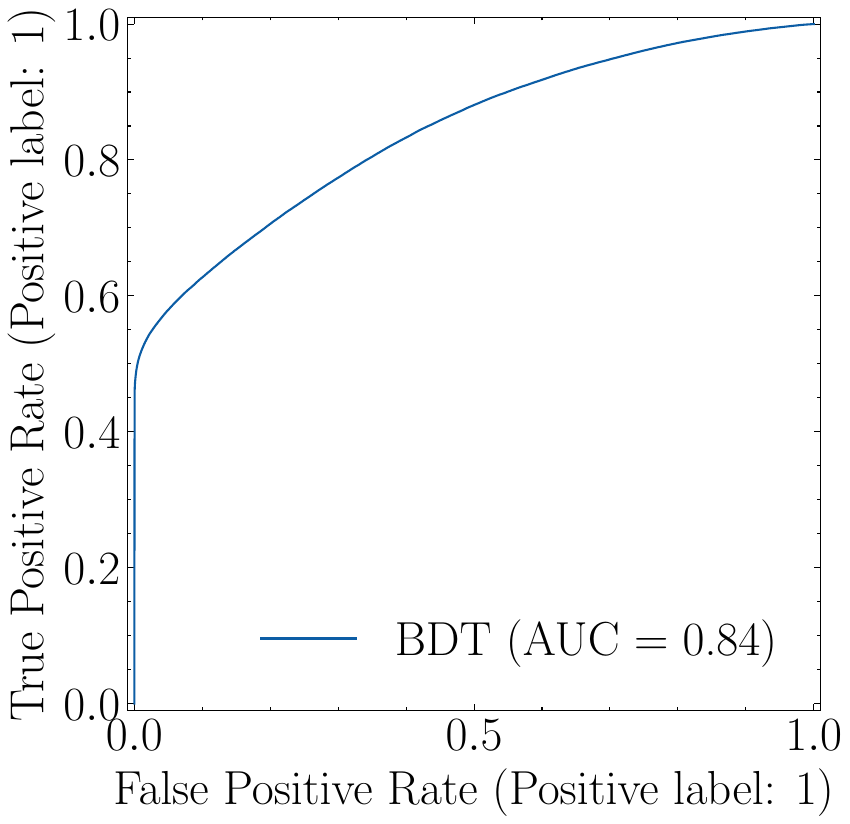}
    \caption{ROC Curve and the corresponding AUC for the final BDT.}
    \label{fig:bdt-roc}
\end{figure}

The final BDT has a ROC AUC of $0.84$ in the test set (similar value was obtained on the validation set during the hyperparameter optimisation loop), as can be observed in Fig.~\ref{fig:bdt-roc}. In this figure, we can also see that the BDT provides a regime with very small false positive rate (contamination) while keeping a relatively large true positive rate (efficiency). The trade-off between the two and how this affects the bounds on the fraction of DM composed of PBH is thoroughly discussed in Section~\ref{sec:constraints}.

\section{BDT Outputs and ML Parameters\label{app:bdt-vs-ml-parameters}}

\begin{figure*}
    \centering
    \includegraphics[width=0.325\linewidth]{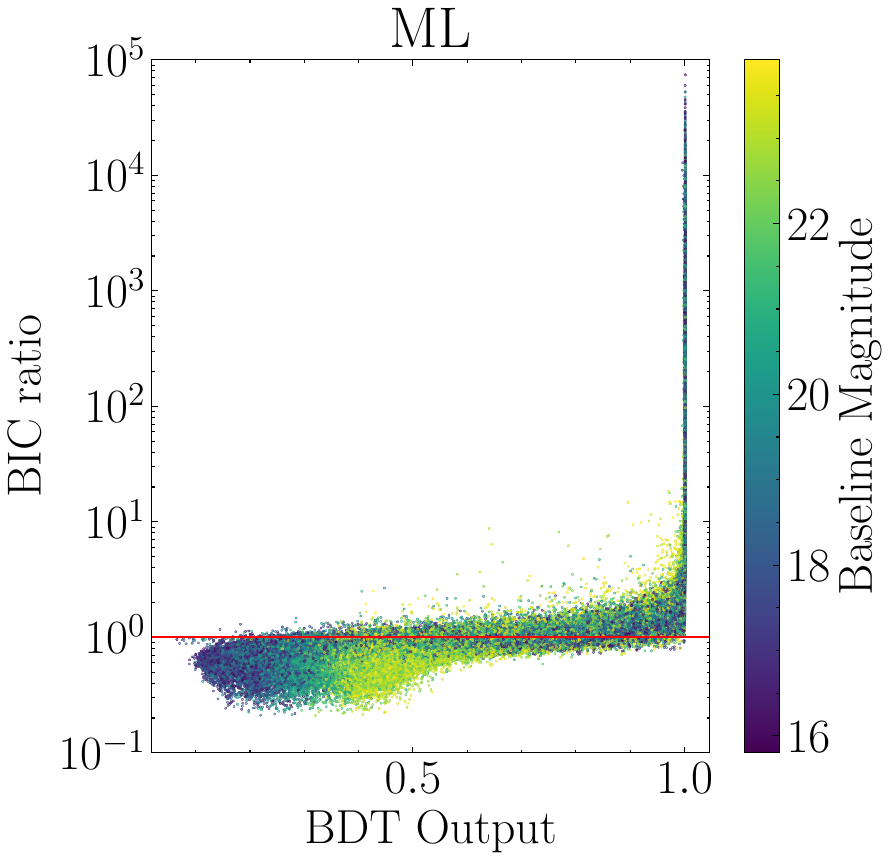}
    \includegraphics[width=0.325\linewidth]{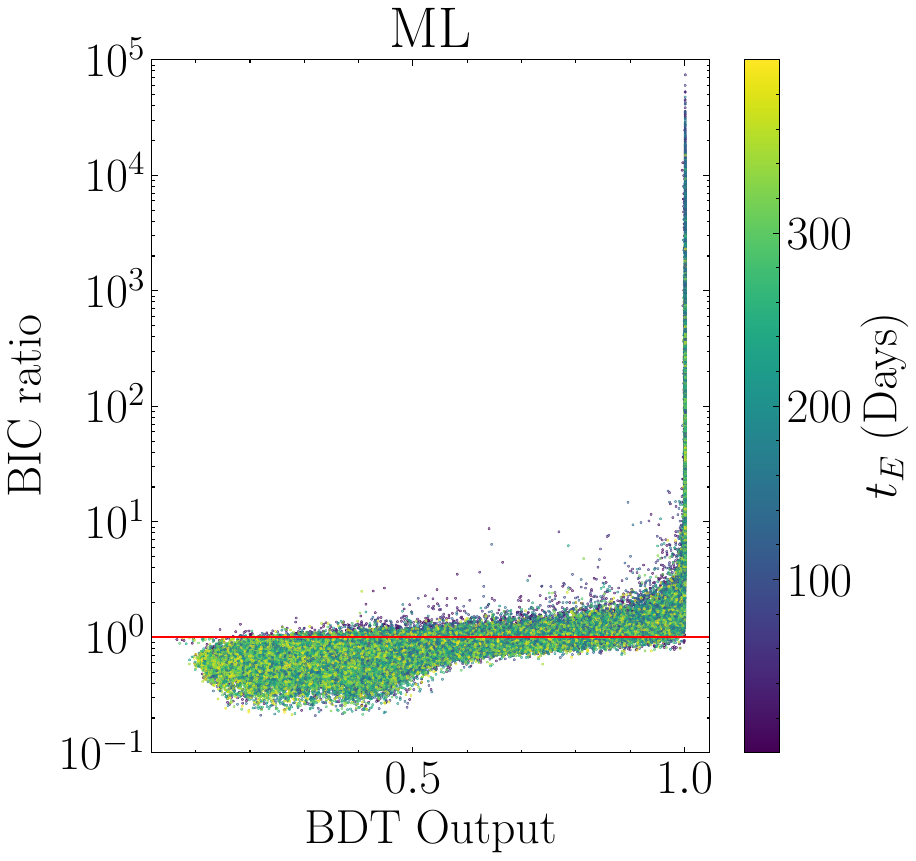}
    \includegraphics[width=0.325\linewidth]{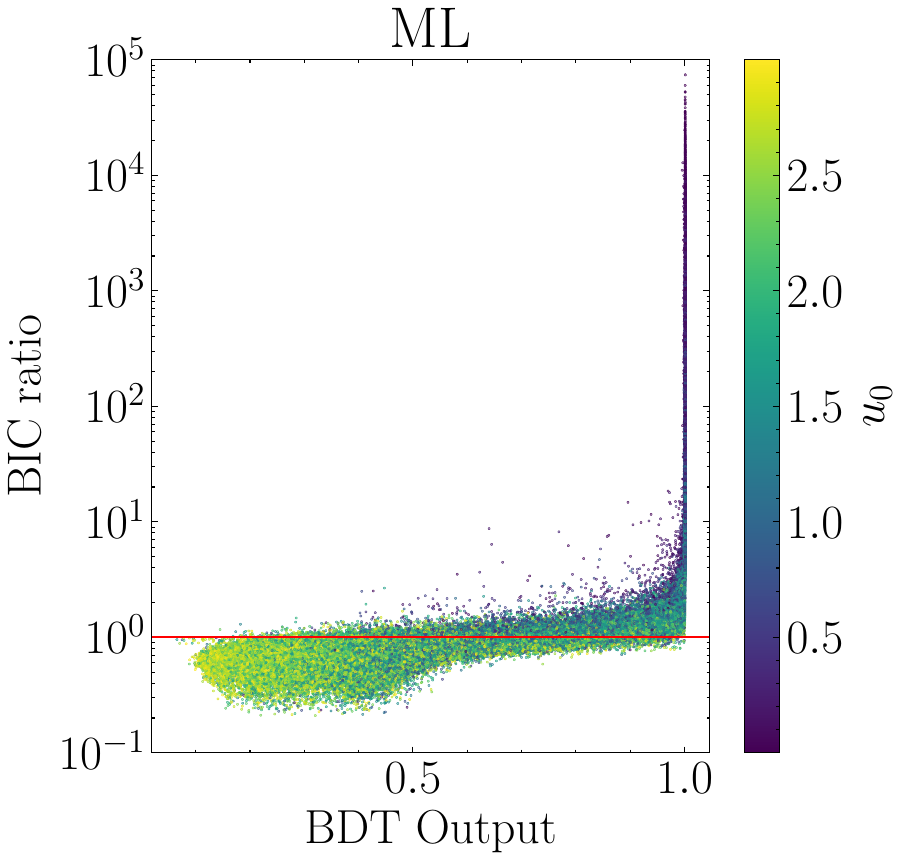}
    \caption{BDT outputs and BIC ratios for the ML class, colour coded by the three microlensing parameters: source baseline magnitude (left), Einstein crossing time (middle), and minimal impact parameter (right).}
    \label{fig:bdt-bic-colour-coded}
\end{figure*}

In Fig.~\ref{fig:bdt-bic-colour-coded} we show how the BIC ratio and BDT output are affected by the microlensing parameters. On the left panel, the scatter points colour indicate the source baseline magnitude, where higher numbers correspond to dimmer stars. We observe that the central mode, where the BDT output is just below $0.5$, the magnitude is the highest, suggesting that the task of differentiating between Constant and ML light curves is particularly difficult in the low brightness regime where the signal to noise ratio is smaller.
In the middle panel, the main observation is there is no discernable impact on the BDT performance over different values of $\tE$. This, however, is a purely visualisation statement, and in Section~\ref{sec:constraints} we discuss the BDT sensitivity for different values of $\tE$.
Finally, on the right panel we observe how both the BDT is especially capable of identifying ML events where the minimal impact parameter is smaller than $1.0$, as this correspond to accentuated lensing magnification. Nonetheless, the BDT is very capable of identifying ML signals for relatively high minimal impact parameter, i.e. $u_0 \gtrsim 1.5 $, that traditional cuts could have missed.

\bibliographystyle{JHEP}
\bibliography{refs}
\end{document}

%% file: universalnewcommands.tex
\newcommand{\acro}[1]{\textsc{\MakeLowercase{#1}}} 

\renewcommand{\tilde}{\widetilde} 

\newcommand{\beq}{\begin{equation}}
\newcommand{\eeq}{\end{equation}}
\newcommand{\bea}{\begin{eqnarray}}
\newcommand{\eea}{\end{eqnarray}}
\newcommand{\nn}{\nonumber}

\usepackage{pdfbase}[2017/03/16]
\usepackage{xparse,ocgbase}
\usepackage{xcolor,calc}
\usepackage{tikzpagenodes,linegoal}
\usetikzlibrary{calc}
\usepackage{tcolorbox}

\ExplSyntaxOn
\let\tpPdfLink\pbs_pdflink:nn
\let\tpPdfAnnot\pbs_pdfannot:nnnn\let\tpPdfLastAnn\pbs_pdflastann:
\let\tpAppendToFields\pbs_appendtofields:n
\def\tpPdfXform{\pbs_pdfxform:nnnnn{1}{1}{}{}}
\let\tpPdfLastXform\pbs_pdflastxform:
\let\cListSet\clist_set:Nn\let\cListItem\clist_item:Nn
\ExplSyntaxOff

\usepackage{pdfbase}[2017/03/16]
\usepackage{xparse,ocgbase}
\usepackage{xcolor,calc}
\usepackage{tikzpagenodes,linegoal}
\usetikzlibrary{calc}
\usepackage{tcolorbox}

\ExplSyntaxOn
\let\tpPdfLink\pbs_pdflink:nn
\let\tpPdfAnnot\pbs_pdfannot:nnnn\let\tpPdfLastAnn\pbs_pdflastann:
\let\tpAppendToFields\pbs_appendtofields:n
\def\tpPdfXform{\pbs_pdfxform:nnnnn{1}{1}{}{}}
\let\tpPdfLastXform\pbs_pdflastxform:
\let\cListSet\clist_set:Nn\let\cListItem\clist_item:Nn
\ExplSyntaxOff

\makeatletter
\NewDocumentCommand{\tooltip}{%
  ssssO{\ifdefined\@linkcolor\@linkcolor\else blue\fi}mO{yellow!20}mO{0pt,0pt}%
}{{%
  \leavevmode%
  \IfBooleanT{#2}{%
    \ocgbase@new@ocg{tipOCG.\thetcnt}{%
      /Print<</PrintState/OFF>>/Export<</ExportState/OFF>>%
    }{false}%
    \xdef\tpTipOcg{\ocgbase@last@ocg}%
    \ocgbase@add@ocg@to@radiobtn@grp{tool@tips}{\ocgbase@last@ocg}%
  }%
  \tpPdfLink{%
    \IfBooleanTF{#4}{%
      /Subtype/Link/Border[0 0 0]/A <</S/SetOCGState/State [/Toggle \tpTipOcg]>>
    }{%
      /Subtype/Screen%
      /AA<<%
        \IfBooleanTF{#3}{%
          /E<</S/SetOCGState/State [/Toggle \tpTipOcg]>>%
        }{%
          \IfBooleanTF{#2}{%
            /E<</S/SetOCGState/State [/ON \tpTipOcg]>>%
            /X<</S/SetOCGState/State [/OFF \tpTipOcg]>>%
          }{
            \IfBooleanTF{#1}{%
              /E<</S/JavaScript/JS(%
                var fd=this.getField('tip.\thetcnt');%
                if(typeof(click\thetcnt)=='undefined'){%
                  var click\thetcnt=false;%
                  var fdor\thetcnt=fd.rect;var dragging\thetcnt=false;%
                }%
                if(fd.display==display.hidden){%
                  fd.delay=true;fd.display=display.visible;fd.delay=false;%
                }else{%
                  if(!click\thetcnt&&!dragging\thetcnt){fd.display=display.hidden;}%
                  if(!dragging\thetcnt){click\thetcnt=false;}%
                }%
                this.dirty=false;%
              )>>%
            }{%
              /E<</S/JavaScript/JS(%
                var fd=this.getField('tip.\thetcnt');%
                if(typeof(click\thetcnt)=='undefined'){%
                  var click\thetcnt=false;%
                  var fdor\thetcnt=fd.rect;var dragging\thetcnt=false;%
                }%
                if(fd.display==display.hidden){%
                  fd.delay=true;fd.display=display.visible;fd.delay=false;%
                }%
               this.dirty=false;%
              )>>%
              /X<</S/JavaScript/JS(%
                if(!click\thetcnt&&!dragging\thetcnt){fd.display=display.hidden;}%
                if(!dragging\thetcnt){click\thetcnt=false;}%
                this.dirty=false;%
              )>>%
            }%
            /U<</S/JavaScript/JS(click\thetcnt=true;this.dirty=false;)>>%
            /PC<</S/JavaScript/JS (%
              var fd=this.getField('tip.\thetcnt');%
              try{fd.rect=fdor\thetcnt;}catch(e){}%
              fd.display=display.hidden;this.dirty=false;%
            )>>%
            /PO<</S/JavaScript/JS(this.dirty=false;)>>%
          }%
        }%
      >>%
    }%
  }{{\color{#5}#6}}%
  \sbox\tiptext{%
    \IfBooleanT{#2}{%
      \ocgbase@oc@bdc{\tpTipOcg}\ocgbase@open@stack@push{\tpTipOcg}}%
    \tcbox[colframe=black,colback=#7,size=fbox,arc=1ex,sharp corners=southwest]{#8}%
    \IfBooleanT{#2}{\ocgbase@oc@emc\ocgbase@open@stack@pop\tpNull}%
  }%
  \cListSet\tpOffsets{#9}%
  \edef\twd{\the\wd\tiptext}%
  \edef\tht{\the\ht\tiptext}%
  \edef\tdp{\the\dp\tiptext}%
  \tipshift=0pt%
  \IfBooleanTF{#2}{%
    \setlength\whatsleft{\linegoal}%
  }{%
    \measureremainder{\whatsleft}%
  }%
  \ifdim\whatsleft<\dimexpr\twd+\cListItem\tpOffsets{1}\relax%
    \setlength\tipshift{\whatsleft-\twd-\cListItem\tpOffsets{1}}\fi%
  \IfBooleanF{#2}{\tpPdfXform{\tiptext}}%
  \raisebox{\heightof{#6}+\tdp+\cListItem\tpOffsets{2}}[0pt][0pt]{%
    \makebox[0pt][l]{\hspace{\dimexpr\tipshift+\cListItem\tpOffsets{1}\relax}%
    \IfBooleanTF{#2}{\usebox{\tiptext}}{%
      \tpPdfAnnot{\twd}{\tht}{\tdp}{%
        /Subtype/Widget/FT/Btn/T (tip.\thetcnt)%
        /AP<</N \tpPdfLastXform>>%
        /MK<</TP 1/I \tpPdfLastXform/IF<</S/A/FB true/A [0.0 0.0]>>>>%
        /Ff 65536/F 3%
        /AA <<%
          /U <<%
            /S/JavaScript/JS(%
              var fd=event.target;%
              var mX=this.mouseX;var mY=this.mouseY;%
              var drag=function(){%
                var nX=this.mouseX;var nY=this.mouseY;%
                var dX=nX-mX;var dY=nY-mY;%
                var fdr=fd.rect;%
                fdr[0]+=dX;fdr[1]+=dY;fdr[2]+=dX;fdr[3]+=dY;%
                fd.rect=fdr;mX=nX;mY=nY;%
              };%
              if(!dragging\thetcnt){%
                dragging\thetcnt=true;Int=app.setInterval("drag()",1);%
              }%
              else{app.clearInterval(Int);dragging\thetcnt=false;}%
              this.dirty=false;%
            )%
          >>%
        >>%
      }%
      \tpAppendToFields{\tpPdfLastAnn}%
    }%
  }}%
  \stepcounter{tcnt}%
}}
\makeatother
\newsavebox\tiptext\newcounter{tcnt}
\newlength{\whatsleft}\newlength{\tipshift}
\newcommand{\measureremainder}[1]{%
  \begin{tikzpicture}[overlay,remember picture]
    \path let \p0 = (0,0), \p1 = (current page.east) in
      [/utils/exec={\pgfmathsetlength#1{\x1-\x0}\global#1=#1}];
  \end{tikzpicture}%
}